\documentstyle[12pt,epsf]{article}

\begin{document}

\baselineskip 6mm
\renewcommand{\thefootnote}{\fnsymbol{footnote}}


\newcommand{\nc}{\newcommand}
\newcommand{\rnc}{\renewcommand}


\rnc{\baselinestretch}{1.24}    
\setlength{\jot}{6pt}       
\rnc{\arraystretch}{1.24}   

\makeatletter
\rnc{\theequation}{\thesection.\arabic{equation}}
\@addtoreset{equation}{section}
\makeatother



\nc{\be}{\begin{equation}}

\nc{\ee}{\end{equation}}

\nc{\bea}{\begin{eqnarray}}

\nc{\eea}{\end{eqnarray}}

\nc{\xx}{\nonumber\\}

\nc{\ct}{\cite}

\nc{\la}{\label}

\nc{\eq}[1]{(\ref{#1})}

\nc{\newcaption}[1]{\centerline{\parbox{6in}{\caption{#1}}}}

\nc{\fig}[3]{

\begin{figure}
\centerline{\epsfxsize=#1\epsfbox{#2.eps}}
\newcaption{#3. \label{#2}}
\end{figure}
}


\def\CA{{\cal A}}
\def\CC{{\cal C}}
\def\CD{{\cal D}}
\def\CE{{\cal E}}
\def\CF{{\cal F}}
\def\CG{{\cal G}}
\def\CH{{\cal H}}
\def\CK{{\cal K}}
\def\CL{{\cal L}}
\def\CM{{\cal M}}
\def\CN{{\cal N}}
\def\CO{{\cal O}}
\def\CP{{\cal P}}
\def\CR{{\cal R}}
\def\CS{{\cal S}}
\def\CU{{\cal U}}
\def\CW{{\cal W}}
\def\CY{{\cal Y}}


\def\IR{{\hbox{{\rm I}\kern-.2em\hbox{\rm R}}}}
\def\IB{{\hbox{{\rm I}\kern-.2em\hbox{\rm B}}}}
\def\IN{{\hbox{{\rm I}\kern-.2em\hbox{\rm N}}}}
\def\IC{\,\,{\hbox{{\rm I}\kern-.59em\hbox{\bf C}}}}
\def\IZ{{\hbox{{\rm Z}\kern-.4em\hbox{\rm Z}}}}
\def\IP{{\hbox{{\rm I}\kern-.2em\hbox{\rm P}}}}
\def\IH{{\hbox{{\rm I}\kern-.4em\hbox{\rm H}}}}
\def\ID{{\hbox{{\rm I}\kern-.2em\hbox{\rm D}}}}


\def\a{\alpha}
\def\b{\beta}
\def\ga{\gamma}
\def\d{\delta}
\def\ep{\epsilon}
\def\ph{\phi}
\def\k{\kappa}
\def\l{\lambda}
\def\m{\mu}
\def\n{\nu}
\def\th{\theta}
\def\rh{\rho}
\def\s{\sigma}
\def\t{\tau}
\def\w{\omega}
\def\G{\Gamma}


\def\half{\frac{1}{2}}
\def\dint#1#2{\int\limits_{#1}^{#2}}
\def\goto{\rightarrow}
\def\para{\parallel}
\def\brac#1{\langle #1 \rangle}
\def\grad{\nabla}
\def\curl{\nabla\times}
\def\div{\nabla\cdot}
\def\p{\partial}
\def\e{\epsilon_0}


\def\Tr{{\rm Tr}\,}
\def\det{{\rm det}}


\def\vare{\varepsilon}
\def\barz{\bar{z}}
\def\barw{\bar{w}}


\def\ad{\dot{a}}
\def\bd{\dot{b}}
\def\cd{\dot{c}}
\def\dd{\dot{d}}
\def\so{SO(4)}
\def\sop{SO(4)^\prime}
\def\bc{{\bf C}}
\def\bfz{{\bf Z}}
\def\bz{\bar{z}}

\begin{titlepage}


\hfill\parbox{3.7cm} {SNUTP 04-003 \\
{\tt hep-th/0402002}}

\vspace{25mm}

\begin{center}
{\Large \bf Exact Seiberg-Witten Map and
Induced Gravity from Noncommutativity}

\vspace{15mm}
Hyun Seok Yang \footnote{Present address: 
Institut f\"ur Physik, Humboldt Universit\"at zu Berlin, Newtonstra\ss e 15,
D-12489 Berlin, Germany; E-mail: hsyang@physik.hu-berlin.de}
\\[10mm]

{\sl School of Physics, Seoul National University,
Seoul 151-747, Korea} \\
\end{center}

\thispagestyle{empty}

\vskip2cm


\centerline{\bf ABSTRACT}
\vskip 4mm
\noindent

We find a closed form for Seiberg-Witten (SW) map between ordinary and
noncommutative (NC) Dirac-Born-Infeld actions. We show that NC Maxwell 
action after the exact SW map can be regarded as ordinary Maxwell action 
coupling to a metric deformed by gauge fields. 
We also show that reversed procedure by inverse SW map leads to a similar
interpretation in terms of induced NC geometry. This implies that
noncommutativity in field theory can be interpreted as field dependent 
fluctuations of spacetime geometry, which genuinely realizes an interesting 
idea recently observed by Rivelles.
\\
\\
Keywords: Noncommutative field theory; Exact Seiberg-Witten map;
Dirac-Born-Infeld action \\
\\
PACS numbers: 11.10.Nx, 11.15.-q, 11.30.-j

\vspace{1cm}

\today

\end{titlepage}

\renewcommand{\thefootnote}{\arabic{footnote}}
\setcounter{footnote}{0}

\section{Dirac-Born-Infeld Action}

We revisit here the equivalence between noncommutative (NC) and
ordinary gauge theories discussed in \ct{sw}. We leave the geometry
of spacetime background fixed and concentrate, instead, on the
dynamics of open string sectors of the theory. To be specific, we
consider open strings attached on $Dp$-branes in
flat spacetime, with metric $g_{\mu\nu}$, in the presence of
a constant Neveu-Schwarz $B$-field.\footnote{Here we will take $g_{\mu\nu}$
with either Lorentz or Euclidean signature since the signature is
inconsequential in our discussions. Also we take $B_{\mu\nu}$ with rank $r=p+1$,
$\mu, \nu = 0, 1, \cdots, p$, for simplicity, although our discussion well 
applies to the case $r < p+1$.} We define a parameter
describing the size of a string as
\be \la{kappa}
\kappa \equiv 2 \pi \alpha^\prime,
\ee
which is a useful expansion parameter in low energy effective
action of D-branes. The worldsheet action is
\bea \la{string-action}
S &=& \frac{1}{2\kappa}\int_{\Sigma} d^2 \sigma (g_{\mu\nu} \p_a
x^\mu \p^a x^\nu - i \kappa B_{\mu\nu} \varepsilon^{ab} \p_a
x^\mu \p_b x^\nu ) - i \int_{\p \Sigma} d\tau A_\mu (x) \p_\tau x^\mu
\nonumber \\
&=& \frac{1}{2\kappa}\int_{\Sigma} d^2 \sigma g_{\mu\nu} \p_a
x^\mu \p^a x^\nu + i \int_{\p \Sigma} d\tau \Bigl( \half B_{\mu\nu} x^\nu
- A_\mu (x) \Bigr) \p_\tau x^\mu,
\eea
where string worldsheet $\Sigma$ is the upper half plane parameterized by $-\infty
\leq \tau \leq \infty$ and $0 \leq \sigma \leq \pi$ and $\p
\Sigma$ is its boundary. The propagator evaluated at boundary
points \ct{sw} is
\begin{equation}\label{open-propagator}
    \langle x^\mu (\tau)  x^\nu (\tau^\prime) \rangle = -
    \frac{\kappa}{2\pi} \Bigl(\frac{1}{G}\Bigr)^{\mu\nu} \log(\tau - \tau^\prime)^2 +
    \frac{i}{2}  \theta^{\mu\nu} \epsilon(\tau - \tau^\prime)
\end{equation}
where $\epsilon(\tau)$ is the step function. Here
\bea \la{open-g-inverse}
&& \left(\frac{1}{G}\right)^{\mu\nu} = \left( \frac{1}{g + \kappa B}
g \frac{1}{g - \kappa B}\right)^{\mu\nu}, \\
\la{open-g}
&& G_{\mu\nu} = g_{\mu\nu} - \kappa^2 (B g^{-1}B)_{\mu\nu}, \\
\la{open-theta}
&& \theta^{\mu\nu} = - \kappa^2 \left( \frac{1}{g + \kappa B}B
\frac{1}{g - \kappa B}\right)^{\mu\nu}.
\eea
From Eqs. \eq{open-g-inverse} and \eq{open-theta},
we have the following relation
\begin{equation} \label{op-cl}
\frac{1}{G} + \frac{\theta}{\kappa} =  \frac{1}{g + \kappa B}.
\ee
The object $G_{\mu\nu}$ has a simple interpretation as the
effective metric seen by the open strings while $g_{\mu\nu}$ is
the closed string metric. Furthermore the coefficient
$\theta^{\mu\nu}$ has a simple interpretation as
\begin{equation}\label{nc-space}
    [ x^\mu (\tau),  x^\nu (\tau) ] = i \theta^{\mu\nu}.
\end{equation}
That is, $x^\mu$ are coordinates on a NC space with
noncommutativity parameter $\theta$ \ct{nc-space1,nc-space2,nc-space3,nc-space4,review}.

For a slowly varying approximation of neglecting
derivative terms, i.e., $ \sqrt{\kappa} |\frac{\p F}{F}| \ll 1$,
the open string effective action on a D-brane was shown to be
given by the Dirac-Born-Infeld (DBI) action \ct{dbi1,dbi2}. Seiberg and Witten,
however, showed \ct{sw} that an explicit form of the effective action
depends on the regularization scheme of two dimensional field
theory defined by the worldsheet action \eq{string-action}, 
which is related to field redefinitions in spacetime.

A sigma model path integral with Pauli-Villars regularization
preserves the ordinary gauge symmetry of open string gauge fields.
With such a regularization, the effective action of a D-brane can
depend on $B$ and $F=dA$ only in the combination $F+B$, since
there is a symmetry $ A \to A + \Lambda, \; B \to B-d\Lambda $, for any
one-form $\Lambda$. In this case, the spacetime low energy effective
action on a single $Dp$-brane is given by the DBI action
\begin{equation}\label{dbi-c}
S(g_s, g, A, B) = \frac{2\pi}{g_s (2\pi \kappa)^{\frac{p+1}{2}}}\int d^{p+1} x
\sqrt{-\det(g + \kappa (F + B))},
\end{equation}
where
\begin{equation}\label{c-f}
F_{\mu\nu} = \p_\mu A_\nu -  \p_\nu A_\mu.
\end{equation}
Note that the effective action is expressed in terms of closed
string variables $g_{\mu\nu}, B_{\mu\nu}$ and $g_s$.

With a point-splitting regularization \ct{sw}, the spacetime effective
action is expressed in terms of NC gauge fields and has the NC
gauge symmetry on the
NC spacetime defined by Eq. \eq{nc-space}.
In this description, the analog of Eq. \eq{dbi-c} is
\begin{equation}\label{dbi-nc}
\widehat{S}(G_s, G, \widehat{A}, \theta) =
\frac{2\pi}{G_s (2\pi \kappa)^{\frac{p+1}{2}}}\int d^{p+1} x
\sqrt{-\det(G + \kappa \widehat{F})}.
\end{equation}
The action depends on the open string variables $G_{\mu\nu}, \theta_{\mu\nu}$
and $G_s$, where the $\theta$-dependence is entirely in the
$\star$ product in the field strength $\widehat{F}$:
\be \la{nc-f}
\widehat{F}_{\mu\nu} = \p_\mu \widehat{A}_\nu -  \p_\nu \widehat{A}_\mu
- i \widehat{A}_\mu \star \widehat{A}_\nu + i \widehat{A}_\nu
\star \widehat{A}_\mu.
\ee
The DBI action \eq{dbi-nc} is definitely invariant under
\begin{equation}\label{nc-gt}
\widehat{\delta}_{\widehat{\lambda}} \widehat{A}_\mu =
\widehat{D}_\mu \star \widehat{\lambda} = \p_\mu \widehat{\lambda} - i
\widehat{A}_\mu \star \widehat{\lambda} + i \widehat{\lambda}
\star \widehat{A}_\mu.
\end{equation}

The ambiguity related to the choice of regularization scheme is a
well-known field redefinition ambiguity present in the effective
action reconstructed from S-matrix. Thus the two descriptions
with different regularizations should be related by a spacetime
field redefinition. Indeed, Seiberg and Witten found a
transformation from ordinary to NC gauge fields in a way that
preserves the gauge equivalence relation between ordinary and NC
gauge symmetries \ct{sw}. The Seiberg-Witten (SW) map relating the gauge
potentials and field
tensors to the first order in $\theta$ is given by
\begin{eqnarray}\label{sw-a}
&& \widehat{A}_\mu = A_\mu - \frac{1}{2}\theta^{\alpha\beta} A_\alpha
(\partial_\beta A_\mu + F_{\beta\mu}) + \CO(\theta^2), \\
\label{sw-f}
&& \widehat{F}_{\mu \nu} = F_{\mu\nu} + \theta^{\alpha\beta}
(F_{\mu\alpha} F_{\nu \beta} - A_\alpha \partial_\beta
F_{\mu\nu}) + \CO(\theta^2).
\eea
Since the commutative and NC descriptions arise from the same open
string theory depending on different regularizations and the
physics should not depend on the regularization scheme,
Seiberg and Witten \ct{sw} argued that\footnote{\label{constant-f}As
already pointed out in \ct{sw}, the comparison cannot be made for
constant $F$ since terms in Eq. \eq{sw-f} such as the form $A \partial F$
contribute in the analysis. It is necessary to integrate by parts
in comparing the DBI actions, and one cannot naively treat $F$ as
a constant.}
\begin{equation}\label{equiv-dbi}
    \widehat{S}(G_s, G, \widehat{A}, \theta) = S(g_s, g, A, B) +
    {\cal O}(\sqrt{\kappa} \partial F).
\end{equation}

The equivalence \eq{equiv-dbi} may also be understood by 
different path integral prescriptions for open strings
ending on a D-brane \ct{tlee,andreev}. If $B$ field is constant, the term involving
the $B$ field in the action \eq{string-action}
can be treated as a part of kinetic term or as a part of boundary
interaction, since it is quadratic in string variables. In the
former case we get the NC DBI action \eq{dbi-nc} and in the
latter case the ordinary one \eq{dbi-c}. Since the two are
obtained by evaluating the same Polyakov string path integral, it
establishes that the NC DBI action is equivalent to the ordinary
one.

First of all, the equivalence \eq{equiv-dbi} determines the open
string coupling constant $G_s$ by demanding that for $F=\widehat{F}=0$
the constant terms in the actions using the two set of variables
are the same:
\begin{equation}\label{Gs-gs}
G_s = g_s \sqrt{\frac{\det G}{\det (g + \kappa B)}}.
\end{equation}
In the comparison \eq{equiv-dbi},
the action $\widehat{S}(G_s, G, \widehat{A}, \theta)$
is expressed in terms of open string parameters while $S(g_s,g, A,
B)$ is in terms of closed string parameters. For an explicit
comparison, we will use the same string variables
for two different descriptions. First, we reexpress Eq. (\ref{dbi-c}) in terms of open
string variables using the conversion relations, Eqs. \eq{op-cl} and
\eq{Gs-gs}, between open and closed string parameters
\begin{equation}\label{op-dbi-c}
S(G_s, G, A, \theta) = \frac{2\pi}{G_s (2\pi \kappa)^{\frac{p+1}{2}}}\int d^{p+1} x
\sqrt{-\det{(G+ F \theta G + \kappa F)}}.
\end{equation}
In what follows, we will often use the matrix notation
\begin{equation}\label{def-matrix}
    AB =  A_{\mu \alpha} B^{\alpha \mu},
    \quad (AB)_{\mu\nu} =  A_{\mu \alpha} {B^\alpha}_\nu, \quad
    \rm{etc}.
\end{equation}
Later we will also consider the equivalence in terms of closed
string variables.

As was explained in \ct{sw}, there is a general description with an arbitrary $\theta$
associated with a suitable regularization that interpolates
between Pauli-Villars and point-splitting. This freedom is
basically coming from the fact that the gauge invariant
combination of $B$ and $F$ in open
string theory is ${\cal F} = B+F$. Thus there is a symmetry of shift in $B$
keeping fixed $B+F$. Given such a symmetry, we may split the $B$
field into two parts and put one in kinetic part and the
rest in boundary interaction part. By taking the background to be $B$ or $B^\prime$,
we should get a NC description with appropriate $\theta$ or
$\theta^\prime$, and different $\widehat{F}$'s.
Hence we can write down a differential equation that describes how
$\widehat{A}(\theta)$ and $\widehat{F}(\theta)$ should change when
$\theta$ is varied, to describe equivalent physics \ct{sw}:
\begin{eqnarray}\label{sw-de-a}
 &&   \delta\widehat{A}_\mu (\theta) =
    -\frac{1}{4}\delta\theta^{\alpha\beta} \Bigl( \widehat{A}_\alpha \star
    (\partial_\beta \widehat{A}_\mu + \widehat{F}_{\beta\mu} )
+ (\partial_\beta \widehat{A}_\mu + \widehat{F}_{\beta\mu}) \star
\widehat{A}_\alpha \Bigr), \\
\label{sw-de-f}
&&   \delta\widehat{F}_{\mu\nu} (\theta) =
    \frac{1}{4}\delta\theta^{\alpha\beta} \Bigl( 2\widehat{F}_{\mu\alpha} \star
    \widehat{F}_{\nu\beta} + 2\widehat{F}_{\nu\beta} \star
    \widehat{F}_{\mu\alpha} -
    \widehat{A}_\alpha \star (\widehat{D}_\beta \widehat{F}_{\mu \nu}
    + \p_\beta \widehat{F}_{\mu\nu}) \nonumber \\
&& \hspace{2.7cm} - (\widehat{D}_\beta \widehat{F}_{\mu \nu}
    + \p_\beta \widehat{F}_{\mu\nu}) \star \widehat{A}_\alpha
    \Bigr).
\end{eqnarray}
Incidentally Eq. \eq{sw-a} and Eq. \eq{sw-f} are a solution of
the differential equations \eq{sw-de-a} and \eq{sw-de-f} to first
order in $\theta$, respectively. An exact solution of the
differential equation \eq{sw-de-f} in the Abelian case
was given in \ct{matrix-sw1,matrix-sw2,matrix-sw3,matrix-sw4}. 
Especially, for the case of rank one gauge field with constant $\widehat{F}$, 
the equation \eq{sw-de-f} can be easily solved to be
\begin{equation}\label{exact-f}
    \widehat{F} = \frac{1}{1 + F\theta} F.
\end{equation}

The freedom in the description just explained above is
parameterized by a two-form $\Phi$ from the point of view of 
NC geometry on the D-brane worldvolume. In this case the
change of variables found by Seiberg and Witten \ct{sw} is given by
\bea \la{op-cl-gen}
&& \frac{1}{{\cal G} + \kappa \Phi} + \frac{\vartheta}{\kappa}
=  \frac{1}{g + \kappa B}, \\
&& \label{Gs-gs-gen}
{\cal G}_s = g_s \sqrt{\frac{\det ({\cal G} + \kappa \Phi)}{\det (g + \kappa B)}}.
\eea
The effective action in these variables are modified to
\begin{equation}\label{dbi-gen}
\widehat{S}_\Phi({\cal G}_s, {\cal G}, \widehat{{\cal A}}, \vartheta) =
\frac{2\pi}{{\cal G}_s (2\pi \kappa)^{\frac{p+1}{2}}}\int d^{p+1} x
\sqrt{-\det({\cal G} + \kappa (\widehat{{\cal F}}+ \Phi))}.
\end{equation}
For every background characterized by $B, g_{\mu\nu}$ and $g_s$,
we thus have a continuum of descriptions labelled by a choice of
$\Phi$. Indeed, for $\Phi=B$ where ${\cal G}=g, \; {\cal G}_s = g_s$
and $\vartheta =0$, $\widehat{S}_\Phi$ recovers the commutative description
\eq{dbi-c} while $\Phi=0$ is the NC description
by \eq{dbi-nc}. So we end up with the most general form of the
equivalence for slowly varying fields, i.e.,
$ \sqrt{\kappa} |\frac{\p F}{F}| $:
\begin{equation}\label{equiv-dbi-gen}
 \widehat{S}_\Phi({\cal G}_s, {\cal G}, \widehat{{\cal A}}, \vartheta) 
= S(g_s, g, A, B) +   {\cal O}(\sqrt{\kappa} \partial F),
\end{equation}
which was proved by Seiberg and Witten \ct{sw} using the change
of variables, \eq{op-cl-gen} and \eq{Gs-gs-gen}, and the
differential equation \eq{sw-de-f}. Using the change of variables
\eq{op-cl-gen} and \eq{Gs-gs-gen}, we also get the
analogue of Eq. \eq{op-dbi-c}
\begin{equation}\label{op-dbi-gen}
S({\cal G}_s, {\cal G}, A, \vartheta; \Phi) =
\frac{2\pi}{{\cal G}_s (2\pi \kappa)^{\frac{p+1}{2}}}\int d^{p+1} x
\sqrt{-\det{({\cal G}+ F \vartheta {\cal G}
+ \kappa (\Phi + F \vartheta \Phi + F))}}.
\end{equation}
Note that the commutative action \eq{op-dbi-gen} is exactly the same as
the DBI action obtained from the worldsheet sigma model 
using $\zeta$-function regularization scheme \ct{andreev}.

\section{Exact Seiberg-Witten Map and Induced Gravity}

In this section we will discuss the meaning of the equivalence
\eq{equiv-dbi-gen} from the field theory point of view. 
The equivalence between the action
\eq{dbi-c}, expressed in the form \eq{op-dbi-gen}, and the action
\eq{dbi-gen} immediately leads to \footnote{The equivalence \eq{sw-equiv}
was also proved in \ct{schupp} in the framework of deformation quantization.
We thank P. Schupp for drawing our attention to their paper.}
\begin{eqnarray}\label{sw-equiv}
 &&  \int d^{p+1} x \sqrt{-\det({\cal G} 
+ \kappa (\widehat{{\cal F}}+ \Phi))} \nonumber \\
 && \hspace{2cm} = \int d^{p+1} x \sqrt{\det{(1+ F \vartheta})}
 \sqrt{-\det{({\cal G} + \kappa (\Phi + {\bf F}))}} 
+  {\cal O}(\sqrt{\kappa}\partial F),
\end{eqnarray}
where
\begin{equation}\label{f-f}
    {\bf F}_{\mu\nu}(x) = \left(\frac{1}{1 + F\vartheta} F\right)_{\mu\nu}(x).
\end{equation}

What is the meaning of the equivalence \eq{sw-equiv} ? First note
that the left hand side of Eq. \eq{sw-equiv} is the NC
description preserving NC gauge symmetry and
$\vartheta$ appears only in the $\star$ product of the field
strength $\widehat{{\cal F}}_{\mu\nu}$. On the other hand,
the right hand side of Eq. \eq{sw-equiv} is the commutative
description preserving ordinary gauge symmetry and
$\vartheta$ explicitly appears in the commutative DBI action.
Next we see from the argument in section 4.1 in \ct{sw} that
Eq. \eq{sw-equiv} is consistent with the differential
equation \eq{sw-de-f} defining the map, e.g., Eqs. \eq{sw-a} and \eq{sw-f},
between ordinary and NC gauge fields. Therefore we
regard the right hand side of Eq. \eq{sw-equiv} as the
SW map of the left hand side, i.e., NC DBI
action labelled by the two-form $\Phi$, valid for every value of the
parameters. We will illustrate this assertion for the case $\Phi = 0$
and for $p=3$, for definiteness, although our following argument
also goes through for general cases. Note that
Eq. \eq{sw-equiv} in the $\Phi = B$ case where $\vartheta = 0$ is
a trivial identity since both sides are equally commutative
descriptions.

For the case $\Phi = 0$ and $p=3$, the identity \eq{sw-equiv} reduces to
\be \label{equiv-phi=0}
\int d^{4} x \sqrt{-\det(G + \kappa \widehat{F})}
    = \int d^{4} X \sqrt{\det{(1+ F \theta)}}
    \sqrt{-\det{(G+ \kappa {\bf F})}},
\ee
where we intentionally distinguished the commutative coordinates $X$ 
from the NC ones for the following discussion.
One can expand both sides of Eq. \eq{equiv-phi=0} in powers of $\kappa$. 
${\cal O}(1)$ implies that there is a measure change 
between NC and commutative descriptions 
\be \label{sw-measure}
d^{4} x = d^{4} X \sqrt{\det{(1+ F \theta)}}.
\ee
In other words, the coordinate transformations, $x^\mu \to X^\mu(x)$, 
between NC and commutative descriptions depend on the dynamical gauge fields.
Since the identity \eq{equiv-phi=0} must be true for arbitrary small $\kappa$,
substituting Eq. \eq{sw-measure} into Eq. \eq{equiv-phi=0} leads to the
following relation 
\be \la{sw-ff}
\widehat{F}_{\mu\nu}(x) = \left( \frac{1}{1 + F\theta} F \right)_{\mu\nu} (X).
\end{equation}

Now ${\cal O}(\kappa^2)$ in Eq. \eq{equiv-phi=0} leads to a remarkable identity
\bea \label{sw-4d}
&& - \frac{1}{4 g_{YM}^2} \int d^4 x \sqrt{-\det{G}}
G^{\mu \alpha} G^{\nu\beta} \widehat{F}_{\mu\nu}
\star   \widehat{F}_{\alpha\beta} \xx
&=& - \frac{1}{4 g_{YM}^2}  \int d^4 x \sqrt{-\det{G}}
\sqrt{\det{(1+ F \theta)}} G^{\mu \alpha} G^{\nu\beta} {\bf F}_{\mu\nu}
{\bf F}_{\alpha\beta} + {\cal O}(\sqrt{\kappa}\partial F),
\eea
where we used the same symbol $x$ again for both descriptions since the
distinction is no longer necessary.
Here we further discuss the ambiguity for constant
$\widehat{F}$ mentioned in footnote \ref{constant-f}.
For the constant $\widehat{F}$ whose exact map is given by Eq.
\eq{exact-f}, the identity \eq{sw-4d} is not quite true since the
right hand side contains the additional factor $\sqrt{\det{(1+ F
\theta)}}$. We see from Eq. \eq{sw-measure} that 
the Jacobian factor for the coordinate transformation, $ x \to X(x)$, precisely 
reproduces the additional factor.

We argued that Eq. (\ref{sw-4d}) defines the exact nonlinear action
of SW deformed electrodynamics. Note that the
identity (\ref{sw-4d}) holds for an arbitrary constant
open string metric $G_{\mu\nu}$. For simplicity, 
we may take $G_{\mu\nu} = \eta_{\mu\nu}$, i.e., flat Minkowski spacetime.
Eq. (\ref{sw-4d}) then takes an interesting form
\begin{equation}\label{sw-flat}
- \frac{1}{4g_{YM}^2} \int d^4 x \widehat{F}_{\mu\nu}
\star   \widehat{F}^{\mu\nu} =
\frac{1}{4g_{YM}^2} \int d^4 x\sqrt{\det{(1+ F\theta)}}
\Bigl(\frac{1}{1 + F\theta} F \frac{1}{1 + F\theta} F \Bigr).
\end{equation}
If we introduce an ``effective non-symmetric metric'' induced by
dynamical gauge fields such that
\begin{equation}\label{ind-metric-flat}
    {\rm g}_{\mu\nu} = \eta_{\mu\nu} + (F\theta)_{\mu\nu},
    \qquad ({\rm g}^{-1})^{\mu\nu} \equiv {\rm g}^{\mu\nu} =  
    \Bigl(\frac{1}{1 + F\theta}\Bigr)^{\mu\nu},
\end{equation}
the NC Maxwell action after the SW map formally looks like 
ordinary Maxwell theory coupled to the effective metric $\mathrm{g}_{\mu\nu}$:
\begin{equation}\label{maxwell-induced}
S = - \frac{1}{4 g_{YM}^2}  \int d^4 x \sqrt{-\det{{\rm g}}} \;
{\rm g}^{\mu \alpha} {\rm g}^{\beta\nu} F_{\mu\nu}
F_{\alpha\beta}.
\end{equation}
It is easy to derive the exact equation of motion from the action \eq{sw-flat}
or \eq{maxwell-induced}
\bea \la{eom-exact}
&& \p_\mu \Biggl[ \sqrt{-{\rm g}} \biggl\{ (\theta {\rm g}^{-1})^{\mu\alpha}
\Tr ({\rm g}^{-1} F {\rm g}^{-1} F) - 2 \Bigl ( (\theta{\rm
g}^{-1}F {\rm g}^{-1} F {\rm g}^{-1})^{\mu\alpha} - (\theta{\rm
g}^{-1}F {\rm g}^{-1} F {\rm g}^{-1})^{\alpha\mu}\Bigr) \xx
&& \hspace{1.7cm} + 2 \Bigl( ({\rm g}^{-1}F {\rm g}^{-1})^{\mu\alpha}
- ({\rm g}^{-1}F {\rm g}^{-1})^{\alpha\mu} \Bigr) \biggr\}
\Biggr] = 0.
\eea
Recently Rivelles observed \ct{rivelles} that the action for NC field theories 
after SW map can be regarded as an ordinary field theory
coupling to a field dependent gravitational background. Our result
genuinely realizes his intriguing idea. The linearized
gravitational coupling of the action \eq{maxwell-induced} exactly reproduces
the result, Eq. (17), in \ct{rivelles}.
It should be remarked that the gravitational field in the action \eq{maxwell-induced}
cannot be interpreted just as a fixed background since it depends
on the dynamical gauge fields.

Now we will show that our result in Eq. \eq{sw-flat} is consistent
with the results in \ct{berrino} where it was proved that the terms of order
$n$ in $\theta$ in the NC Maxwell action via SW map form a
homogeneous polynomial of degree $n+2$ in $F$ ({\it Proposition
3.1}) and explicitly presented the deformed action up to order $\theta^2$.
It is obvious that Eq. \eq{sw-flat} satisfies their {\it Proposition
3.1}. The explicit form of the $\theta$-expanded action in \ct{berrino}
has the following expression using the matrix notation
\eq{def-matrix}\footnote{To avoid a confusion, we point out that
$F^2=F_{\mu\nu}F^{\mu\nu}$ in \ct{berrino} corresponds to our
$-F^2=F_{\mu\nu}F^{\nu\mu}$ in the matrix notation \eq{def-matrix}.}
\begin{eqnarray}\label{italian-action}
    S &=& \frac{1}{4 g_{YM}^2} \int d^4 x \Tr \Bigl( (1+ \half \Tr F\theta)
    F^2 -2 F\theta F^2 + F\theta F^2 \theta F + 2 F \theta F \theta F^2 \nonumber \\
    && \qquad - \Tr(F \theta) F \theta F^2 + \frac{1}{8} (\Tr F\theta)^2
    F^2 - \frac{1}{4} \Tr(F \theta)^2 F^2 + {\cal O}(\theta^3)
    \Bigr).
\end{eqnarray}
It is straightforward to reproduce the result \eq{italian-action}
from Eq. \eq{sw-flat} using the formulas
\bea
&& \sqrt{\det{(1+ F \theta)}} = 1 + \half \Tr F \theta -\frac{1}{4}
\Tr (F \theta)^2 + \frac{1}{8} (\Tr F \theta)^2 + {\cal
O}(\theta^3), \xx
&& \frac{1}{1+ F \theta} = 1 - F \theta + (F \theta)^2
+ {\cal O}(\theta^3). \nonumber
\eea

Another interesting case arises from the choice $\Phi_{\mu\nu} = - B_{\mu\nu}$,
which naturally appears in Matrix models \ct{sw,seiberg}. In this
case, using the metric $g_{\mu\nu}$ with Euclidean
signature instead,
\begin{equation}\label{matrix}
    \theta = \frac{1}{B}, \qquad G= - \kappa^2 B \frac{1}{g} B,
    \qquad G_s = g_s \sqrt{\det(\kappa B g^{-1})}
\end{equation}
and
\begin{equation}\label{matrix-f}
    (\widehat{F} + \Phi)_{\mu\nu} = i B_{\mu \lambda} [X^\lambda, X^\sigma]_\star B_{\sigma
    \nu},
\end{equation}
where
\begin{equation}\label{X}
    X^\mu = x^\mu + \theta^{\mu\nu} \widehat{A}_\nu.
\end{equation}
The DBI action related to Matrix models has a more natural
description, the so-called background independent formulation, in
terms of closed string variables \ct{seiberg}.
Using the relations \eq{matrix} and \eq{matrix-f},
the equivalence \eq{equiv-dbi-gen} can be recast as
\bea \label{sw-matrix}
&& \frac{2\pi}{g_s (2\pi)^{\frac{p+1}{2}}} \int \frac{d^{p+1} x}{{\rm |Pf \theta|}}
\sqrt{\det(\delta^\nu_\mu - \frac{i}{\kappa}g_{\mu\lambda}[X^\lambda,
X^\nu]_\star)} \xx
&& \quad = \frac{2\pi}{g_s (2\pi \kappa)^{\frac{p+1}{2}}}\int d^{p+1} x
\sqrt{\det(1 - \widehat{F}\theta )} \sqrt{\det(g + \kappa (B + \widehat{{\bf
F}}))}, \xx
&& \quad = \frac{2\pi}{g_s (2\pi \kappa)^{\frac{p+1}{2}}}\int d^{p+1} x
\sqrt{\det(g + \kappa (B + F))},
\eea
where
\begin{equation}\label{matrix-ncf}
    \widehat{{\bf F}}_{\mu\nu}(x) = \left(\frac{1}{1-\widehat{F}\theta} 
\widehat{F} \right)_{\mu\nu}(x).
\end{equation}
We point out that the identity in Eq. \eq{sw-matrix} is come out
from the equivalence \eq{equiv-dbi-gen} by expressing the
NC DBI action \eq{dbi-gen} in terms of closed string variables. 
One can check that our result \eq{sw-matrix} for slowly varying fields 
is consistent with the exact SW map obtained by completely independent 
way in \ct{matrix-sw1,matrix-sw2,matrix-sw3,matrix-sw4}.

As was discussed in \ct{seiberg}, $X^\mu$ are background
independent, i.e., $\theta$-independent, coordinates and can be used to
describe the coordinates on D-branes for all values of $\theta$.
Thus one can see that both sides of Eq. \eq{sw-matrix} are
background independent since the integral measure
$\frac{1}{(2\pi)^{\frac{p+1}{2}}} \int \frac{d^{p+1} x}{{\rm |Pf
\theta|}}$, which is a trace over the Hilbert space of the algebra
\eq{nc-space} \ct{review}, and $F+B$ are background independent objects.
Similarly to Eq. \eq{sw-equiv}, Eq. \eq{sw-matrix} also defines a map
between NC (Matrix) and commutative descriptions, but now in terms
of inverse SW map.

Note that
\begin{equation}\label{matrix-bf}
 B  + \widehat{{\bf F}} = \frac{1}{1 - \widehat{F}\theta} B.
\end{equation}
In the zero slope limit, $\kappa \to 0$, now keeping $g_{\mu\nu}$
and $g_{YM}^2$ fixed, Eq. \eq{sw-matrix} gives rise to an intriguing identity
\begin{equation}\label{sw-matrix-zero}
\frac{1}{4g_{YM}^2}\int d^{p+1} x {\cal F}_{\mu\nu} {\cal F}^{\mu\nu} =
\frac{1}{4g_{YM}^2}\int d^{p+1} x \sqrt{\det{\widehat{{\rm g}}}} \;
\widehat{{\rm g}}^{\mu \alpha} \widehat{{\rm g}}^{\beta\nu} B_{\mu\nu}
B_{\alpha\beta},
\end{equation}
where ${\cal F} = B + F$ and
\be \la{matrix-metric}
\widehat{{\rm g}}_{\mu \nu} = \delta_{\mu\nu} - (\widehat{F} \theta)_{\mu\nu}, 
\qquad \widehat{{\rm g}}^{\mu \nu} = \Bigl(\frac{1}{1 - \widehat{F} 
\theta}\Bigr)^{\mu\nu}.
\ee
A naive interpretation of the identity \eq{sw-matrix-zero} may be
that fluctuations $F$ with respect to the background $B$ induce fluctuations 
of (NC) geometry from the Matrix model side. It will be very interesting 
if one can understand this picture from $D$-brane perspective.

\section{Discussion}

Our results in the present paper may have many interesting
implications in both string theory and field theory. In particular
we showed that the dual description via SW map describes a
fluctuating geometry induced by gauge fields and noncommutativity,
in a sense, reflects the presence of a fluctuating ``medium". 
The spacetime geometry (nonsymmetric gravity \ct{moffat1,moffat2,moffat3})
determined by the metric \eq{ind-metric-flat} does deserve further study.
The geometry to the leading order in $\theta$ was studied in
\ct{rivelles} where it was shown that the plane wave solution
in \ct{jackiw-et-al,cai,abe-et-al} corresponds to a geodesic motion 
of massless particles in that gravitational field.

There are several interesting open issues for the future:
Non-Abelian generalization \ct{na-dbi1,na-dbi2,na-dbi3}, an exact SW map 
for DBI actions with derivative corrections 
\ct{der-correction1,der-correction2,der-correction3}, and
to find an exact SW map for currents by incorporating
matter fields \ct{matrix-sw3,matrix-sw4,our}. 
It will surely be interesting to reexamine noncommutative $U(1)$ instantons 
\ct{instanton1,instanton2,instanton3,instanton4,instanton5} in view of
the action \eq{maxwell-induced}, which may be helpful to ponder
topological issues in the commutative description via SW map.
We will report our progress on these issues elsewhere in the near
future.

{\bf Notes added}   Several points raised in this paper have been clarified
since the original version of this paper was posted to the archive.

Our method to obtain the exact SW map is extremely simple. In particular, we
got the SW maps for the measure change and the field strength,
Eq. \eq{sw-measure} and Eq. \eq{sw-ff}, respectively. It was shown in
\ct{banerjee-yang} that these maps can be derived from the equivalence between
the star products $\star_\omega$ and $\star_B$ defined by the symplectic forms 
$\omega = B + F$ and $B$, respectively, in the context of deformation
quantization and one can find the exact SW map for an adjoint scalar field 
using the results. It was also shown there that topological invariants in NC
gauge theory are mapped to the usual Chern classes via the exact SW map.

The closed form for the exact SW map of NC electrodynamics, 
Eq. \eq{maxwell-induced}, turned out to pose an important physics 
about emergent gravity \ct{emergent1,emergent2,hsyang}. It was shown 
in \ct{emergent1,emergent2} that NC $U(1)$ instantons are equivalent 
to gravitational instantons via the exact SW map, indeed posed at the last of
the Discussion above and speculated at the last paragraph in section 6 
of \ct{banerjee-yang}. In particular, we showed in \ct{hsyang} that 
self-dual electromagnetism in noncommutative spacetime is equivalent 
to self-dual Einstein gravity.

Using the exact SW maps presented in this paper and \ct{banerjee-yang},
Mukherjee and Saha showed \ct{mukherjee-saha1,mukherjee-saha2} that 
either NC Chern-Simons or NC Maxwell-Chern-Simons model with scalar matter 
in the adjoint representation and without any potential term does not have any 
nontrivial BPS soliton in the sector which has a smooth commutative limit. 
Their results clearly show that, in these models, there is no non-trivial, 
non-perturbative solution depending on the NC parameter and vanishing smoothly 
along with it, which is consistent with the topological property 
of soliton solutions.

It was also discussed in \ct{kersting-ma} that NC field theory interpreted as ordinary
field theory embedded in a gravitational background induced by gauge fields  
is quite similar to quantum field theory coupling to a non-symmetric metric
background at the phenomenological level and the two theories predict 
experimentally measurable consequences in the high energy process
such as the pair annihilation $e^+ e^- \to \gamma \gamma$. 
The geometry generated by NC gauge fields was further studied in \ct{muth} for
the massive Klein-Gordon field.

{\bf Acknowledgment}   We thank Roman Jackiw for an inspiring question
stimulating this work and Rabin Banerjee and Choonkyu Lee for
helpful discussions. This work was supported by the Brain Korea 21
Project in 2003.

\newpage


\end{document}